\documentclass[manuscript,screen]{acmart}
\AtBeginDocument{%
  \providecommand\BibTeX{{%
    \normalfont B\kern-0.5em{\scshape i\kern-0.25em b}\kern-0.8em\TeX}}}

\setcopyright{acmcopyright}
\copyrightyear{2021}
\acmYear{2021}
\acmDOI{}

\acmConference[DTRAP '21]{DRAP '21: Digital Threats: Reasearch And Practice}{}{}
\acmBooktitle{DRAP '21: Digital Threats: Reasearch And Practice}
\acmPrice{0.00}
\acmISBN{}

\begin{document}

\title{Active and Passive Collection of SSH key material for cyber threat intelligence}

\author{Alexandre Dulaunoy} \affiliation{
\institution{CIRCL} \streetaddress{16 Boulevard d'Avranches}
\city{Luxembourg} \country{Luxembourg}} \email{firstname.name@circl.lu}

\author{Jean-Louis Huynen} \affiliation{
  \institution{CIRCL} \streetaddress{16 Boulevard d'Avranches}
  \city{Luxembourg} \country{Luxembourg}} \email{firstname.name@circl.lu}

\author{Aurelien Thirion} \affiliation{
  \institution{CIRCL} \streetaddress{16 Boulevard d'Avranches}
  \city{Luxembourg} \country{Luxembourg}} \email{firstname.name@circl.lu}

\renewcommand{\shortauthors}{Dulaunoy, et al.}

\begin{abstract}


This paper describes a system for storing historical forensic artifacts
collected from SSH connections. This system exposes a REST API in a similar
fashion as passive DNS databases, malware hash registries, and SSL notaries with
the goal of supporting incident investigations and monitoring of infrastructure.

\end{abstract}



\maketitle

\section{Introduction}

CSIRTs operate passive DNS \cite{draft-dulaunoy-dnsop-passive-dns-cof-07} and
passive SSL \cite{dulaunoy_alexandre_passive_2015} databases to support incident
response. These historical data foster infrastructure attribution and threat
intelligence at large. In this field note, we describe a passive SSH~\cite{pssh}
the we developed and discuss how it constitutes a worthy addition to the CSIRTS'
toolbox.

SSH, and especially OpenSSH implementation, is the main tool for remote
management as it offers secure channels for accessing servers' shell, moving
data, and tunneling other protocols. As OpenSSH is present on a lot of servers,
MacOS and recent Windows computer, as well as IoT devices it makes it very
appealing for attackers as vector of attack and means of command and control
(see FriztFrog\cite{fritzfrog} for instance).

\section{Fingerprinting SSH protocol}

The SSH protocol\cite{rfc4253} provides a significant number of fingerprints to
track similar infrastructures as it uses public-key cryptography to authenticate
clients and server. In the frame of infrastructure fingerprinting, the host-key,
that is the key used to authenticate a server to clients is a first data point
one can collect. The common use-case is that servers provide a host-key to the
clients that is persistent between client connections. This ceremony follows a
Trust On First Sight model: the server presents the host-key to the client that,
on the first connection, is prompted to decided whether to trust the key or not.
Once trusted, host-keys are stored in a cache on the client side against which
presented host-keys will be checked on subsequent connections. Server host-keys
are therefore persistent, they are usually generated when setting up ssh and
left untouched afterwards.

Whereas SSH1 only provides algorithm negotiation between the client and the
server for bulk encryption, SSH2 introduces the negotiation for the host-key,
message authentication, hash function, session key exchange and data
compression. These are others data point usable for fingerprinting SSH endpoints.

Salesforce also created HASSH, a network fingerprinting standard which
can be used to identify specific Client and Server SSH implementations (see
\cite{hasshblog} and \cite{hasshcode}).

Fingerprinting SSH servers requires to initiate a TCP three-way handshake with
the server and read the first SSH protocol message (see Figure~
\ref{sshhandshake}\footnote{This image is a modified
  version of one found at https://github.com/salesforce/hassh and will be
  replaced with original content for the final version}).


\begin{center}
  \includegraphics[width=0.8\textwidth]{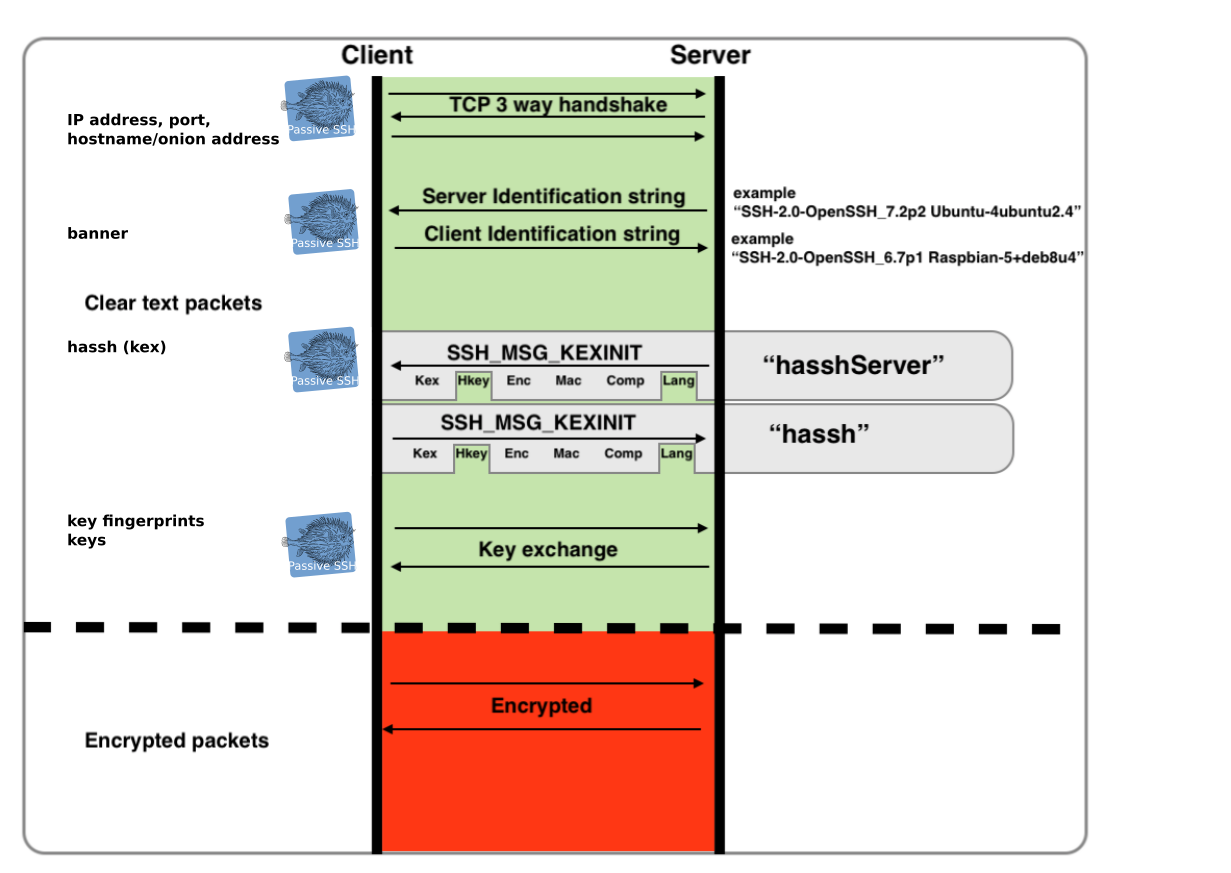}
\captionof{figure}{Fingerprinting SSH first packets}
\label{sshhandshake}
\end{center}


\section{Passive SSH Design and Implementation}


We developed a tool to fingerprint SSH servers on the Internet\footnote{Available here
\url{https://github.com/D4-project/passive-ssh/blob/main/bin/ssh_scan.py}}. The
tool is developed in python and uses paramiko\footnote{\url{https://www.paramiko.org/}} which is an
implementation of SSH2 in python to interact with SSH servers.

The data points collected for fingerprinting each servers are the following:

\begin{itemize}
  \item the remote server's banner that consists of:
    \begin{itemize}
      \item server version / idstring,
      \item supported key exchange algorithms,
      \item supported encryption algorithms,
      \item supported mac algorithms,
      \item supported compression algorithms,
      \item server language.
    \end{itemize}

  \item the remote server keys MD5 fingerprint, base64 representation, and name;
  \item the remote server IP;
  \item eventually we compute the remote server's hassh.
\end{itemize}


In order to store these SSH servers fingerprints, we developed the tool that is
the main focus of this paper: Passive SSH\footnote{Available
at\url{https://github.com/D4-project/passive-ssh}}. The tool is written in
python 3 (released as open source project), persistence is achieve using a
redis-compatible back-end\footnote{Redis is a key/value in-memory back-end that
allows for high read/write throughputs, see \url{https://redis.io/}. Redis is used to allow fast-lookup from numerous applications which require low-latency response.}. Passive SSH
provides a REST API to push data into the datastore and to retrieve signatures.

\section{Statistics}

CIRCL operates a passive SSH instance accessible to FIRST, TF-CSIRT, CNW
members, and vetted researchers. This instance includes an
internet-wide scan of SSH available on IPv4 and TCP port 22.

\begin{table}[ht]
\caption{IPv4 addresses with SSH enabled}
\begin{tabular}{ll}
\hline
Banners             & 99.474   \\
IPv4 addresses      & 10.301.105 \\
Tor Onion addresses & 92
\end{tabular}
\end{table}

\begin{table}[ht]
\caption{SSH key types seen on SSH enabled IPv4 addresses }
\begin{tabular}{ll}
\hline
ssh-ed25519  & 5.271.642  \\
ssh-rsa      & 7.291.675 \\
ssh-dss      & 1.530.867 \\
ecdsa-sha2-nistp256 & 5.078.182 \\
ecdsa-sha2-nistp521 & 42.679 \\
ecdsa-sha2-nistp384 & 5.843 \\
\end{tabular}
\end{table}
\section{Use-Cases}

SSH banners are still very uniform over a large number of scanned IP hosts. This
provides a nifty way to cluster groups of hosts depending of their SSH
implementation installed especially the outliers.

\subsection{Tracking attackers' infrastructure}

A Passive SSH service allows to readily answer valuable questions:
\begin{itemize}
\item Is this host key new to my environment?
\item What server presented this key before?
\item Did this server already presented this key?
\item When was the last time a specific host key was seen in use?
\item How many host key were presented by a single IPv4 address?
\end{itemize}

To make the most of passive SSH, we are currently working on a module for
MISP\cite{wagner2016misp} that will allow to automatically enrich MISP events
with these new data points. Discovering attack infrastructure, as described in
Attribution of Advanced Persistent Treats\cite{steffens2020advanced}, can be
performed by using the fingerprints and hassh in Passive SSH. In addition, key
renewal in SSH is less frequent than TLS which then allows to keep pivoting
information for a longer period.

\subsection{Finding vulnerable equipments}

In the same manner as Gasser et al.~\cite{gasser_deeper_2014}, a passive SSH
database can be used to identify equipments accessible on the Internet running
vulnerable software or using weak cryptographic material. The former is readily
accessible by interrogating Passive SSH's banner endpoint for vulnerable banners
but the latter requires more efforts as some computations may be required to
identify weak cryptographic material and leak keys. For this purpose, we plan on
interfacing another project of ours, snake-oil-crypto
\footnote{https://github.com/D4-project/snake-oil-crypto} to discover weak key
materials.

\subsection{Uncloacking tor}

Deanonymizing tor hidden services through SSH server fingerprinting is a strategy
that has two main limitations: (1) each hidden service is usually kept on
separate onion addresses. This separation of concern prevent leaking the ties
between services, and (2) if a SSH service is also running to administer the
machine from the Internet, its access may be controlled by other layers that
prevents it to be directly accessible (e.g. the use of a Bastion host, foreign
ports, or port-knocking strategies).

We scanned SSH service on port 22 and 2222 on 8194 onion addresses known to be
alive in our AIL framework~\cite{mokaddem2018ail} onion list and found 92 SSH
service. Crossing onion SSH servers with publicly available SSH service on the
Internet, we found 9 of those 92 onions have a corresponding SSH fingerprint in
our passive SSH database.

\section{Future work}

Our passive SSH implementation shows a significant source of network
fingerprints to discover vulnerable infrastructure, adversary network
infrastructure or weak cryptographic materials. We could improve scanning
strategies such as BGP networks announces (IPv6 network addresses) or
certificate transparency logs\cite{laurie2014certificate} to come across
recently exposed network and IoT devices with ephemeral SSH access. SSH banners
include various information which could increase the granularity of the
fingerprints. Improving high-speed SSH handshake identification by using the
Passive SSH database to train a neural network model.

\section{Acknowledgments}
We would like to thank the anonymous reviewers for their useful comments and
suggestions.\\

\includegraphics[width=0.48\textwidth]{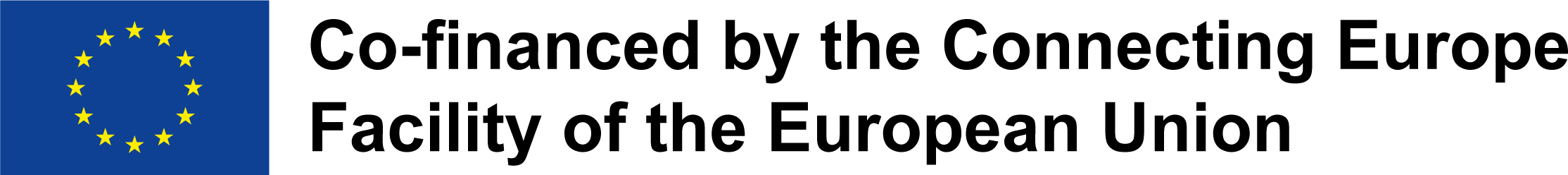}

\bibliographystyle{plain}
\bibliography{referecences/referecences.bib}

\begin{thebibliography}{10}

\bibitem{pssh}
{{CIRCL team}}.
\newblock {{Passive SSH}}, 12 2020.

\bibitem{draft-dulaunoy-dnsop-passive-dns-cof-07}
Alexandre Dulaunoy, Aaron Kaplan, Paul~A. Vixie, and Henry Stern.
\newblock {Passive DNS - Common Output Format}.
\newblock Internet-Draft draft-dulaunoy-dnsop-passive-dns-cof-07, Internet
  Engineering Task Force, June 2020.
\newblock Work in Progress.

\bibitem{dulaunoy_alexandre_passive_2015}
Alexandre Dulaunoy and Eireann Leverett.
\newblock Passive ssl passive - {Detection and Reconnaissance Techniques, to
  Find, Track, and Attribute Vulnerable Devices}.

\bibitem{gasser_deeper_2014}
Oliver Gasser, Ralph Holz, and Georg Carle.
\newblock A deeper understanding of {SSH}: Results from internet-wide scans.
\newblock In {\em 2014 {IEEE} Network Operations and Management Symposium
  ({NOMS})}, pages 1--9. {IEEE}.

\bibitem{fritzfrog}
{G}uardicore.
\newblock Fritzfrog: A new generation of peer-to-peer botnets, 08 2020.

\bibitem{laurie2014certificate}
Ben Laurie.
\newblock Certificate transparency.
\newblock {\em Communications of the ACM}, 57(10):40--46, 2014.

\bibitem{mokaddem2018ail}
Sami Mokaddem, G{\'e}rard Wagener, and Alexandre Dulaunoy.
\newblock Ail-the design and implementation of an analysis information leak
  framework.
\newblock In {\em 2018 IEEE International Conference on Big Data (Big Data)},
  pages 5049--5057. IEEE, 2018.

\bibitem{hasshblog}
Ben Reardon.
\newblock Open sourcing hassh a profiling method for ssh clients and servers,
  09 2018.

\bibitem{hasshcode}
Salesforce.
\newblock "hassh" - a profiling method for ssh clients and servers.

\bibitem{steffens2020advanced}
Timo Steffens.
\newblock Advanced persistent threats.
\newblock In {\em Attribution of Advanced Persistent Threats}, pages 3--21.
  Springer, 2020.

\bibitem{wagner2016misp}
Cynthia Wagner, Alexandre Dulaunoy, G{\'e}rard Wagener, and Andras Iklody.
\newblock Misp: The design and implementation of a collaborative threat
  intelligence sharing platform.
\newblock In {\em Proceedings of the 2016 ACM on Workshop on Information
  Sharing and Collaborative Security}, pages 49--56. ACM, 2016.

\bibitem{rfc4253}
T.~Ylonen and Ed. C.~Lonvick.
\newblock The secure shell (ssh) transport layer protocol.
\newblock RFC 4253, RFC Editor, 10 2006.

\end{thebibliography}
\end{document}